\documentclass[12pt]{spieman}  
\usepackage{amsmath,amsfonts,amssymb}
\usepackage{graphicx}
\usepackage{setspace}
\usepackage{tocloft}
\usepackage{multirow}
\usepackage{tablefootnote}
\usepackage{makecell}
\graphicspath{{Figures/}}

\newcommand{\ed}{\textcolor{black}}

\newcommand{\caltech}{Department of Astronomy, California Institute of Technology, 1200 E. California Blvd., Pasadena, CA 91125, USA}

\newcommand{\ucsc}{Department of Astronomy \& Astrophysics, University of California, Santa Cruz, CA 95064, USA}
\newcommand{\ucsb}{Department of Physics, University of California, Santa Barbara, CA 93106, USA}
\newcommand{\keck}{W. M. Keck Observatory, 65-1120 Mamalahoa Hwy, Kamuela, HI 96743, USA}
\newcommand{\ucla}{Department of Physics \& Astronomy, University of California, 430 Portola Plaza, Los Angeles, CA 90095, USA}
\newcommand{\jpl}{Jet Propulsion Laboratory, California Institute of Technology, 4800 Oak Grove Dr., Pasadena, CA 91109, USA}
\newcommand{\ucsd}{Center for Astrophysics and Space Sciences, University of California, San Diego, La Jolla, CA 92093}

\newcommand{\northwestern}{Center for Interdisciplinary Exploration and Research in Astrophysics (CIERA) and Department of Physics and Astronomy, Northwestern University, Evanston, IL 60208, USA}
\newcommand{\osu}{Department of Astronomy, The Ohio State University, Columbus, OH 43210, USA}
\newcommand{\nice}{Universit\'{e} C\^{o}te d'Azur, Observatoire de la C\^{o}te d'Azur, CNRS, Laboratoire Lagrange, Bd de l'Observatoire, CS 34229, 06304 Nice cedex 4, France}
\newcommand{\grenoble}{Universit\'{e} Grenoble Alpes, CNRS, Institut de Plan\'{e}tologie et d'Astrophysique (IPAG), F-38000 Grenoble, France}

\title{Vortex Fiber Nulling for Exoplanet Observations: Implementation and First Light}

\author[a,*]{Daniel Echeverri}    
\author[a]{Jerry Xuan}
\author[a]{Nemanja Jovanovic}
\author[b]{Garreth Ruane}
\author[c]{Jacques-Robert Delorme}
\author[a,b]{Dimitri Mawet}
\author[b]{Bertrand Mennesson}
\author[b]{Eugene Serabyn}
\author[b]{J. Kent Wallace}
\author[d]{Jason Wang}
\author[e]{Jean-Baptiste Ruffio}
\author[f]{Luke Finnerty}
\author[a]{Yinzi Xin}
\author[g]{Maxwell Millar-Blanchaer}
\author[a]{Ashley Baker}
\author[b]{Randall Bartos}
\author[f]{Benjamin Calvin}
\author[c]{Sylvain Cetre}
\author[c]{Greg Doppmann}
\author[f]{Michael P. Fitzgerald}
\author[g]{Sofia Hillman}
\author[a]{Katelyn Horstman}
\author[d]{Chih-Chun Hsu}
\author[a]{Joshua Liberman}
\author[f]{Ronald Lopez}
\author[h]{Evan Morris}
\author[a]{Jacklyn Pezzato}
\author[i]{Caprice L. Phillips}
\author[j,k]{Bin B. Ren}
\author[e]{Ben Sappey}
\author[a]{Tobias Schofield}
\author[h]{Andrew J. Skemer}
\author[g]{Connor Vancil}
\author[i]{Ji Wang}

\affil[a]{\caltech}
\affil[b]{\jpl}
\affil[c]{\keck}
\affil[d]{\northwestern}
\affil[e]{\ucsd}
\affil[f]{\ucla}
\affil[g]{\ucsb}
\affil[h]{\ucsc}
\affil[i]{\osu}
\affil[j]{\nice}
\affil[k]{\grenoble}

\cftpagenumbersoff{figure}
\cftpagenumbersoff{table} 
\begin{document} 
\maketitle

\begin{center}
    Submitted to Journal of Astronomical Telescopes, Instruments, and Systems. \\
    (Accepted August 30, 2023)
\end{center}

\begin{abstract}
Vortex fiber nulling (VFN) is a single-aperture interferometric technique for detecting and characterizing exoplanets separated from their host star by less than a diffracted beam width. VFN uses a vortex mask and single mode fiber to selectively reject starlight while coupling off-axis planet light with a simple optical design that can be readily implemented on existing direct imaging instruments \ed{that can feed light to an optical fiber}. With its axially symmetric coupling region peaking within the inner working angle of conventional coronagraphs, VFN is more efficient at detecting new companions at small separations than conventional direct imaging, thereby increasing the yield of on-going exoplanet search campaigns. We deployed a VFN mode operating in K band ($2.0{-}2.5~\mu$m) on the Keck Planet Imager and Characterizer (KPIC) instrument at the Keck II Telescope. In this paper we present the instrument design of this first on-sky demonstration of VFN and the results from on-sky commissioning, including planet and star throughput measurements and predicted flux-ratio detection limits for close-in companions. The instrument performance is shown to be sufficient for detecting a companion $10^3$ times fainter than a 5\textsuperscript{th} magnitude host star in 1 hour at a separation of 50~mas (1.1$\lambda/D$). This makes the instrument capable of efficiently detecting substellar companions around young stars. We also discuss several routes for improvement that will reduce the required integration time for a detection by a factor ${>}$3. 
\end{abstract}


{\noindent \footnotesize\textbf{*}Daniel Echeverri,  \linkable{dechever@caltech.edu} }

\begin{spacing}{1}   

%

\section{Introduction}

The National Academies' Astro2020 Decadal Review identified exoplanet science as a key frontier in the coming decade and specifically emphasized development in direct imaging~\cite{Astro2020_Decadal}. Through high spectral resolution characterization of light from the exoplanet itself, direct imaging provides unique insight into exoplanet properties including atmospheric composition\cite{Konopacky2013_MolecularMapping, Ruffio2021_DeepHR8799, Wang2021_KPICScience,Wang2022_KPICCORetrieval,Xuan2022_KPICRetrieval, Wang2023_CORetrivalHR8799}, spin~\cite{Snellen2014_Spin, Bryan2020_Spin, Xuan2020_Spin}, planetary radial velocity\cite{Ruffio2023_KPICExomoons,Wang2021_KPICScience,Delorme2021_KPIC}, and cloud coverage~\cite{Crossfield2014_DopplerImagingDemo}. However, of the more than 5,400 confirmed exoplanets to date, less than 50 have been detected with direct imaging~\cite{Akenson2013_Archive}. A major limitation to direct detection is that young, giant exoplanets amenable to direct imaging seem to be rarer than originally expected at wide separations ($\gtrsim$10~AU or correspondingly $0.1^{\prime\prime}$ for stars within 100 parsec)\cite{Bowler2018_DIOccurrRates}. This has led to a yield of only a few new planets in previous imaging surveys of hundreds of stars~\cite{Nielsen2019_GPIYield, Vigan2021_SphereDemographics}.
By combining Hipparcos~\cite{Perryman1997_Hipparcos, VanLeeuwen2007_HipValidation} and Gaia~\cite{GaiaCollaboration2018_DR2, Gaia2021_eDR3} astrometry to identify and target accelerating stars that show promise for harboring a substellar  companion~\cite{Kervella2019_GaiaAccel,Brandt2021_HipGaiaCatalog,DeRosa2019_AccelChallenges}, the yield of recent direct imaging surveys has been improved~\cite{Currie2023_GaiaAccelDetect, Kuzuhara2022_AccelDetect, Franson2023_Astrometric,DeRosa2023_AccelDetect,Hinkley2023_GravityAccelDetect}.
Nevertheless, these campaigns use coronagraphs which are limited to separations ${\gtrsim}3\lambda/D$ from the star (${\sim}13$ AU at 100~parsec for $\lambda{=}2.2~\mu$m and $D$=10~m), where $\lambda$ is the wavelength and $D$ is the telescope diameter. Radial velocity (RV) surveys\cite{Fulton2021_OccurRates,Rosenthal2021_OccurRates} suggest that the peak in the giant planet population is closer-in to the star than state-of-the-art AO imaging survey instruments  can currently access\cite{Macintosh2014_GPIInstrument,Beuzit2019_SPHEREInstrument}, indicating that the yield can be further improved by unlocking access to these smaller separations.

\ed{Larger telescopes, such as the upcoming Extremely Large Telescopes (ELTs), will target smaller physical separations than their predecessors due to the scaling of $\lambda/D$ but they will still be limited in angular resolution by their coronagraphs. Additionally, observing in short wavelengths enables the detection of planets at smaller physical separations for the same scaling reason but accessing ${\sim}1\lambda/D$ would allow those same planets to be further characterized, after detection, at longer wavelengths without disappearing behind the inner working angle of the coronagraph. Thus, new technologies are still needed to push to ${\sim}1\lambda/D$ though} few methods currently exist. Sparse aperture masking \ed{(SAM) has a long history working in this regime, and has even been deployed on the James Webb Space Telescope. Since SAM does not remove the starlight though, the demonstrated ground-based contrast is currently around 7 magnitudes ($1.6\times10^{-3}$) in L' band (${\sim}3.8~\mu m$) at 80~mas ($1\lambda/D$ in L' on Keck)~\cite{Sallum2019_KeckNRM, Hinkley2011_KeckNRM}.}
Long-baseline multi-aperture interferometers such as the very successful VLTI-GRAVITY can also access this angular separation range~\cite{Abuter2017_GravityFirstLight}; \ed{in 2023, the \mbox{GRAVITY} instrument demonstrated a detection of a companion at a contrast of $8.2\times10^{-5}$ with a separation of 86~mas~\cite{Hinkley2023_GravityAccelDetect}}.
Dark hole digging \ed{with GRAVITY}~\cite{Pourre2022_GravityDarkHole} or long-baseline multi-aperture nullers such as the VLTI-Asgard/NOTT projects are also an option~\cite{Defrere_2015FLL,Defrere-2022LBN}. \ed{Nevertheless, GRAVITY, ASGARD/NOTT, and other such projects} require complex and costly one-of-a-kind infrastructure and thus are not suitable for widespread implementation. 

Vortex Fiber Nulling (VFN) is a single-aperture, interferometric technique that sidesteps \ed{many of} these limitations and can provide access to planets between 0.5 and 2.0$\lambda/D$, or less, in a full 360$^\circ$ region around a host star all at once. It builds on the heritage of previous fiber nulling techniques~\cite{Serabyn2019_PFN} which utilize the spatial and modal filtering properties of a single-mode fiber (SMF) to simplify the optical design of classical nullers. When used in unison with conventional coronagraphs following up on Gaia-Hipparcos accelerators, VFN can improve the separation coverage and bridge the gap between direct imaging and RV and transit surveys\cite{Echeverri2021_BroadbandVFN,Ren2023_VegaImagingAndRV}. VFN also has direct application to the Habitable Worlds Observatory mission recommended by the Astro2020 Decadal Review. The stringent pointing and wavefront control requirements set by the mission's coronagraph instrument, which are on the order of tens of picometers\cite{Habex2020_FinalReport, LUVOIR2019_FinalReport}, far exceed the requirements for VFN\cite{Ruane2019SPIE}. This means that with very few modifications, a VFN mode on the telescope would open a new search area around targets and thereby increase the mission yield.

Initially proposed in 2018\cite{Ruane2018_VFN} and demonstrated in the lab shortly thereafter\cite{Echeverri2019_VFN,Echeverri2020_VFNSPIE,Echeverri2021_BroadbandVFN}, VFN is now operating on-sky at the Keck II Telescope. This paper introduces the first VFN instrument and covers results from the on-sky commissioning phase. Section~\ref{sec:VFN} briefly describes the VFN technique while Section~\ref{sec:KPICVFN} presents the optical design, requirements, and how the instrument operates. Section~\ref{sec:data} reports the observations obtained from April 2022 to January 2023 to determine the on-sky measured star and planet throughput. Section~\ref{sec:analysis} uses these throughput measurements to determine sensitivity and detection limits, showing that VFN is ready for science observations. Finally, Section~\ref{sec:NextSteps} presents the next steps for improving the instrument performance.

\section{The Vortex Fiber Nulling Concept} \label{sec:VFN}
\begin{figure}
    \begin{center}
    \includegraphics[width=\linewidth]{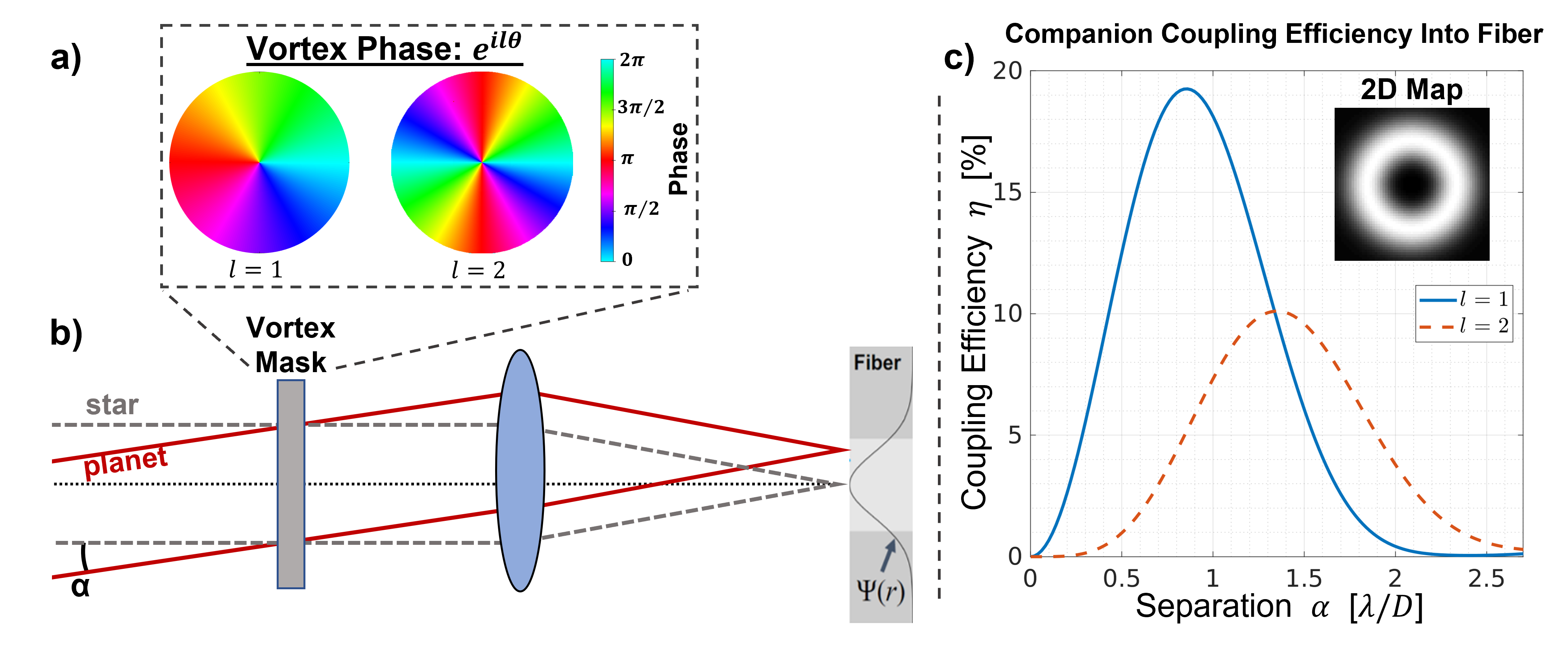}
    \end{center}
    \caption{
    (a)~Azimuthally varying phase pattern introduced by a charge $l=1$ and charge $l=2$ vortex mask. (b)~Diagram of a VFN system with the vortex mask upstream of an SMF in the image plane. The star is aligned with the SMF such that the target planet lands at an off-axis angle, $\alpha$, with respect to the fundamental mode of the fiber, $\Psi(r)$. (c)~Coupling efficiency, $\eta$, versus angular separation from the optical axis, $\alpha$, for a charge 1 (solid blue) and 2 (dashed orange) VFN system, assuming a circular aperture. The inset shows the coupling efficiency for all points in a field-of-view centered on the star/fiber, displaying the axial symmetry in the coupling profile.
    \label{fig:Concept}} 
\end{figure}

The VFN concept places an optical vortex mask upstream of an injection unit that couples light from a point-source into a single-mode fiber~\cite{Ruane2018_VFN}. The vortex imparts an azimuthal phase ramp of the form $\exp{(il\theta)}$, where $l$ is an integer known as the charge and $\theta$ is the azimuthal coordinate~\cite{Swartzlander2001_OVC, Swartzlander2009_OVCTheory}. Figure~\ref{fig:Concept}(a) shows the phase ramp for charge 1 and 2 vortex masks. In this paper, we assume the vortex is in a pupil plane as shown in Fig~\ref{fig:Concept}(b). However, it can also be placed in a focal plane with nearly identical performance. The pupil implementation has the benefit that the ideal F/\# for planet coupling with the vortex is the same as without the vortex. This design allows VFN to be added as a complementary mode to high-contrast imaging instruments with \ed{existing fiber injection units} without changing the system F/\#~\cite{Ruane2019SPIE}. The vortex phase propagates through the system to the SMF plane and is selectively filtered by the fiber. At the fiber plane, the fraction of light, $\eta(r)$, that couples into the fiber's axially symmetric fundamental mode, $\Psi(r)$, for an incident electric field, $E(r,\theta)$, is given by the overlap integral,
\begin{equation} \label{eq:overlap}
    \eta(r) = \left|\int\Psi(r)E(r,\theta)dA\right|^2,
\end{equation}
where $r$ and $\theta$ are radial coordinates centered on the fiber and $E$ and $\Psi$ are normalized by their \ed{individual} total power \ed{such that $\int\left|\Psi(r)\right|^2dA = 1$ and $\int\left|E(r,\theta)\right|^2dA = 1$}. \ed{Thus $\eta$ provides the fraction of total light incident on the fiber plane that couples into the SMF.} Due to the vortex phase, the electric field for a point source aligned to the fiber can be expressed in the form $E(r,\theta)=f(r)\exp{(il\theta)}$. The overlap integral is then separable and has an azimuthal term, $\int_{0}^{2\pi} \exp{(il\theta)}d\theta$, that computes to zero for non-zero integer values of $l$. This results in theoretically perfect rejection of on-axis light and this central region with zero coupling is thus referred to as the ``null". 
For off-axis point sources, the vortex phase is not symmetric over the fiber mode, which results in a non-zero coupling efficiency at small angular separations from the optical axis. An exoplanet can thus be observed by aligning its host star onto the center of the SMF so that the star is rejected while the off-axis planet light couples in and is carried by the fiber to a detector. 

The coupling efficiency for the exoplanet depends on its separation and the charge of the vortex, as shown in Fig.~\ref{fig:Concept}(c). Charge $l{=}1$, the solid blue line, has a peak of $\ed{\eta}{\sim}20$\% at $0.9~\lambda/D$ and $l{=}2$, the dashed orange line, has $\ed{\eta}{\sim}10$\% at $1.4~\lambda/D$. These values assume a circular unobstructed pupil but very similar performance, to within $\pm$1\%, is achieved on the apertures of most major telescopes~\cite{Ruane2019SPIE}. \ed{For reference without the vortex, the maximum coupling on a circular aperture is $\eta{\sim}80$\% at $0~\lambda/D$ (ie. on-axis).} A key benefit of VFN is that the coupling efficiency is axially symmetric and creates a ring, shown in the inset of Fig.~\ref{fig:Concept}(c), where light is transmitted through the fiber. This allows VFN to simultaneously search a complete annular region around a star all at once for new companions\ed{. O}ther SMF-based direct imaging instruments \ed{achieve higher throughput by observing without a vortex but they} require a raster scan around the star or multiple fibers to cover the same area~\cite{Lovis2017_RISTRETTOConcept}.

\ed{VFN can theoretically achieve bandwidths of $\Delta\lambda/\lambda\gtrsim50$\% or more since the null is wavelength independent and the mode-field diameter (MFD) of SMFs, which sets the coupling efficiency, scales roughly with the diffraction in the focal plane. In practice, the bandwidth is limited by the chromaticity of the vortex mask and the transmission of the SMF; current technology can achieve bandwidths commensurate with astronomical bands of ${\sim}$20\%, dominated by leakage in vector vortex masks.}

As with other nulling and coronagraphic techniques, VFN is sensitive to tip-tilt pointing errors and wavefront aberrations. However, the degree to which the null is affected depends on the vortex charge. The null for charge 1 follows a second-order power law for tip-tilt and coma aberrations while charge 2 is only second-order for astigmatism\cite{Ruane2019SPIE}. 
All other aberrations either do not affect the null or have a fourth-order effect such that they are insignificant compared to tip, tilt, coma, and astigmatism~\cite{Ruane2019SPIE}. The bright fringe from VFN is also affected by wavefront aberrations though the main effect for modest aberrations is to slightly distort the ring while leaving the average radial coupling peak relatively unaffected. 
Further implications of wavefront error (WFE) on observations and the achievable on-sky performance are covered in Sections~\ref{sec:analysis} and \ref{sec:NextSteps}.

For more details on the underlying principles behind the VFN concept, we refer readers to the original VFN paper by Ruane et al.~\cite{Ruane2018_VFN} and subsequent works which flesh out the concept, design requirements, and related trades~\cite{Ruane2019SPIE,Echeverri2019b_VFN}. VFN was demonstrated in the laboratory at visible wavelengths with monochromatic nulls~\cite{Echeverri2019_VFN,Echeverri2020_VFNSPIE} of ${\sim}5{\times}10^{-5}$ and polychromatic nulls of ${<}10^{-4}$ with a 15\% bandwidth~\cite{Echeverri2021_BroadbandVFN}.

\section{A VFN Mode for KPIC} \label{sec:KPICVFN}
Given the successful demonstration of VFN in the laboratory and the simplicity with which it can be implemented on existing SMF-fed instruments, we added a VFN demonstrator mode to the Keck Planet Imager and Characterizer (KPIC) instrument at the Keck II Telescope. KPIC is a bridge between the facility Adaptive Optics (AO) system~\cite{Wizinowich2000} and NIRSPEC~\cite{McLean1998_NIRSPEC,McLean2000_NIRSPEC,Martin2018_NIRSPEC}, the existing slit-based high-resolution infrared spectrograph at Keck Observatory. KPIC uses SMFs to spatially filter residual starlight and background while coupling planet light and providing a highly stable linespread function on the detector~\cite{Mawet2017_KPIC}. KPIC has been deployed in phases, with Phase I bringing the core elements of the fiber injection unit needed to couple light into the fiber. This phase was commissioned in 2018~\cite{Delorme2021_KPIC} and allowed KPIC to operate in ``direct spectroscopy" (DS) mode, in which the fiber is aligned directly with the targeted exoplanet. With a resolving power of $R{=}\lambda/\Delta\lambda\sim35,000$ in $K$ band (2.0-2.5~$\mu$m), the KPIC DS mode provided the first high-resolution spectra of HR~8799~c, d and e. This led to the first spin measurements for the planets along with constraints on the planet radial velocity and atmospheric properties~\cite{Wang2021_KPICScience,Wang2023_CORetrivalHR8799}. Robust measurements of atmospheric abundances have also been demonstrated with KPIC DS mode data~\cite{Xuan2022_KPICRetrieval,Wang2022_KPICCORetrieval}. Though the DS mode maximizes coupling efficiency, it requires precise knowledge of the companion position so it is only practical for characterization of known companions rather than for making new detections. Deployed in February 2022, Phase II of KPIC brought several upgrades including a vortex mask to enable VFN~\cite{Echeverri2022_KPICPhaseII}. 
The VFN mode's wider, annular coupling region allows KPIC to search for new exoplanets at small separations and expands the proven capabilities of the DS mode to previously unknown systems. 

\subsection{Optical Layout}
\begin{figure}[b]
    \begin{center}
    \includegraphics[width=\linewidth]{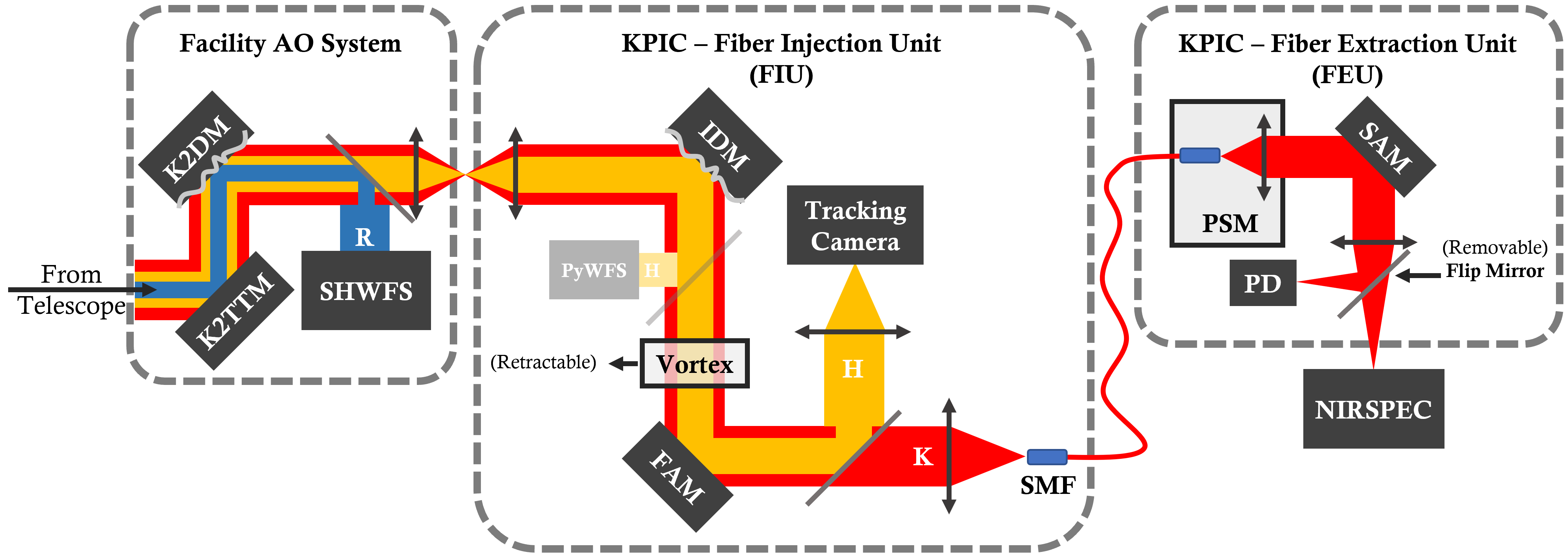}
    \end{center}
    \caption{
    Schematic diagram of the KPIC VFN mode. The facility AO system feeds AO-corrected light to the fiber injection unit (FIU) which includes the vortex and SMF necessary for VFN. The FIU also includes a fiber alignment mirror (FAM) and tracking camera for aligning the stellar PSF with the SMF. The fiber extraction unit (FEU) reimages the output of the SMF and aligns it with the input of NIRSPEC. A flip miror in the FEU optionally picks off the light to go to a dedicated photodetector (PD) rather than NIRSPEC for fast readouts during calibrations. Note that various elements in the optical layout have been omitted for simplicity since they are not relevant to the VFN mode; Jovanovic et al.\cite{Jovanovic2020_KPICPhaseII} and Echeverri et al.\cite{Echeverri2022_KPICPhaseII} provide the layout of the optical path in the Phase II DS mode.
    \label{fig:OpticalLayout}} 
\end{figure} 

For a detailed description of the Phase II instrument design, optical layout, performance, and other non-VFN KPIC modes, we refer readers to the general Phase II instrument paper in preparation by Jovanovic et al. In the meantime, we defer to two other publications which describe most of the Phase II design\cite{Jovanovic2020_KPICPhaseII,Echeverri2022_KPICPhaseII}. We also note that many of the individual elements of the instrument, particularly the tracking system and fiber bundle, remain largely unchanged from the original Phase I design reported by Delorme et al.~\cite{Delorme2021_KPIC} This section covers only the aspects related to the KPIC VFN mode.

Figure~\ref{fig:OpticalLayout} shows the optical layout of KPIC when observing in the new VFN mode. The Facility AO system, composed of a tip-tilt mirror (K2TTM), a 349-actuator deformable mirror (K2DM), and a Shack-Hartmann Wavefront Sensor (SHWFS) operating in the visible (400-950 nm), feeds AO-corrected light to the fiber injection unit (FIU) of KPIC. The first element in the FIU is a dedicated 1000-actuator deformable mirror (IDM) internal to KPIC which is currently used to provide wavefront offsets and non-common-path aberration (NCPA) corrections for the various KPIC modes. An optional dichroic after the IDM can be inserted to send light to a near-infrared pyramid wavefront sensor (PyWFS)~\cite{Bond2020_PyWFS} though that sensor is not currently used for VFN observations; the capabilities unlocked by the PyWFS will be covered in Section~\ref{sec:NextSteps}. A relay (not shown) then sends the light to a translating pupil stage which holds the vortex mask for VFN. 
The system currently uses a narrowband charge 2 vector vortex mask characterized by Mawet et al.~\cite{Mawet2010b_KBandVort} to have a central wavelength of 2.225~$\mu$m such that KPIC VFN observations are currently focused on K band \ed{and are not limited on-sky by the vortex leakage within the ${\sim}$500~nm width of the band.} Another relay (not shown) sends the light to a dedicated fast tip-tilt mirror, called the fiber alignment mirror (FAM), used to steer the star's point spread function (PSF) to the desired position relative to the SMF and control residual jitter from the AO system. The FAM works in closed loop with a tracking camera that receives light short of 1.85~$\mu$m reflected to it by a pickoff dichroic. The tracking camera is a First Light Imaging C-RED2 detector\cite{Gibson2020_CRED2} 
that forms an image with minimal NCPA relative to the final fiber plane. 
Wavelengths longer than 1.85~$\mu$m are transmitted through the dichroic to a triplet lens which focuses the light onto a fiber bundle containing four SMFs whose cores are separated by ${\sim}125$~$\mu$m (${\sim}800$~mas in K-band). These SMFs are identical though the relative coupling efficiency into each varies slightly; for simplicity, we will generally refer to the bundle as ``the SMF" indicating the science fiber within the bundle chosen for maximum coupling on a given day. 

The fiber extraction unit (FEU) collimates the output of the SMF and projects the pupil onto the slit alignment mirror (SAM), which is conjugate to the cold stop inside NIRSPEC. The SMF and collimating lens sit on the pupil-steering mechanism (PSM) which can be translated to ensure the collimated beam is centered on the cold stop. The SAM steers the reimaged PSF from the SMF to align with the NIRSPEC slit. These two actions ensure maximum throughput to the detector, which provides the final spectra. 
A flip mirror can be optionally inserted into the beam just before the NIRSPEC input to send the light to a single-pixel InGaAs photodetector (PD: Thorlabs PDA10DT) which is used for fast readouts during calibrations. Note that various elements in the optical path have been omitted here for simplicity since they are not relevant to the VFN mode or to this work. 
However, the design presented in this section covers all elements relevant to the KPIC VFN mode.

\subsection{Tracking System} \label{sec:tracking}
The tracking system plays a key role since VFN places strict requirements on the positioning accuracy of the star over the center of the fiber to maintain the null.
The requisite accuracy depends on the vortex charge: a charge 2 vortex, like the one currently installed, requires that the star be within $0.3\lambda/D$ ($\lesssim13.5$~mas) on average to maintain a null of ${<}10^{-3}$ but charge 1 requires better than $0.04\lambda/D$ ($\lesssim1.8$~mas) for the same null level\cite{Ruane2019SPIE}. Besides the requirement on average PSF position, the requirement on acceptable root-mean-square (RMS) jitter (standard deviation of the PSF position) is $0.18\lambda/D$ ($\lesssim8$~mas) RMS for charge 2 and $0.03\lambda/D$ ($\lesssim1.3$~mas) RMS for charge 1 to maintain a null ${<}10^{-3}$~\cite{Ruane2019SPIE}.

The facility AO system currently provides sufficiently low jitter residuals, $\sim$6-7~mas RMS, for our charge 2 vortex. However, the average PSF position slowly drifts beyond the requirement when not corrected by the KPIC tracking system. The KPIC tracking camera is used to identify the PSF position and drive the dedicated FAM to compensate for residual pointing errors~\cite{Delorme2021_KPIC}. During daytime calibrations, a two-dimensional (2D) tip-tilt scan is performed with the FAM to scan the PSF over the SMF and identify the camera pixel coordinates that center the beam on the fiber. Once on-sky, these coordinates are used as the ``goal" to which the PSF is driven. \ed{The system has proven to be extremely stable such that the coordinates only drift by a fraction of a pixel over days but we still do the 2D scan before each night to ensure optimal coupling}. The PSF from a vortex has a ring-like intensity profile~\cite{Swartzlander2001_OVC,Kotlyar2007} such that standard centroiding algorithms like center-of-mass, quadratic fitting, and Gaussian fitting, do not generally work without modification. The tracking camera on KPIC operates in H band (1.48-1.70~$\mu$m) though, rather than the design wavelength for the vortex of 2.225~$\mu$m. Thus, the PSF seen by the camera is a combination of the VFN ring plus an Airy PSF arising from the chromatic leakage term in the vector vortex mask~\cite{Ruane2019_scalarVC}. The zero-order leakage from our vortex mask is large enough in H band that the PSF has a clear Airy pattern core on the tracking camera. This allows us to identify the star using a 2D Gaussian fitting algorithm that provides sub-pixel accuracy since the Airy core is oversampled at 4.1~pixels per full-width at half-maximum. Operating in closed loop with the FAM, the KPIC tracking system accurately maintains the average PSF position to within $\sim0.2$~mas. However, the system currently focuses on slow drift only and the jitter residuals remain roughly the same as received from the AO system. Control over the jitter is primarily limited by the tracking loop software and resonances in the system. There is on-going work to improve the control software to operate faster and specifically reduce the jitter residuals.

\subsection{Calibration and Observing Procedure} \label{sec:obsStrat}
Calibrations for the KPIC VFN mode are currently limited by the calibration light source. This \ed{source} is \ed{a} broadband\ed{,} thermal lamp coupled into a SMF to inject light at the input of the facility AO bench and is too faint to measure the VFN null on the PD. Thus the system needs to first be fully aligned and calibrated in the DS mode so that NIRSPEC, which is orders of magnitude more sensitive than the PD and can detect the signal at the VFN null point, can be used to apply corrections for the VFN mode. The DS calibrations involve fiber finding, as described in Sec.~\ref{sec:tracking}, followed by an NCPA correction on the PD to quickly minimize wavefront aberrations at the fiber plane. The DS NCPA correction is performed by scanning each Zernike aberration with varying amplitudes using the IDM to determine where the maximum coupling occurs. The output of the fiber, still in the DS mode, is then aligned to NIRSPEC by scanning the PSM and the SAM to make sure the beam passes unimpeded through the NIRSPEC pupil and slit. With KPIC aligned to maximize the DS mode performance, the vortex is translated into the beam and a smaller set of Zernike aberrations and amplitudes are scanned to minimize the on-axis signal as measured by NIRSPEC. This minimizes the wavefront aberrations in the VFN mode and optimizes the null depth. 

Two additional calibrations are performed on-sky before switching to the target of interest. 
First, a star with many spectral lines and a well-known radial velocity is observed to obtain a wavelength solution for the night. Then an A0 telluric standard star is observed to sample the telluric features in the desired patch of sky. 
With those calibrations complete, the target host star's spectrum is measured by aligning it with the center of the fiber and taking exposures with the vortex out, such that the star is well-coupled into the SMF. This measurement is used in post-processing to fit for the residual starlight on top of the companion signal and as a radiometric calibration of the throughput achieved on the given night. Finally, the vortex is moved into the beam to null the star and observe the companion. 
The background in KPIC spectra is primarily dominated by instrumental background within NIRSPEC due to a light leak in the spectrograph first reported by Lopez et al.\cite{Lopez2020_NIRSPECUpgradeSpecs}. Thus, background spectra are generally taken off-sky at the beginning or end of the night with the same integration time as used on-sky during the night. To account for possible variations in the background level over time, 
a ``nodding" technique can instead be used where the star is bounced between two fibers so that each NIRSPEC frame has a fiber with the target spectrum and another with the background, enabling nod-subtraction. For VFN observations, this is often not needed since the residual coupled starlight due to wavefront aberrations yields a stellar photon noise that is at least an order of magnitude above the background photon noise for stars with a K~band magnitude of 6 or brighter.

\subsection{System Validation} \label{sec:summitVal}

\begin{figure}
    \begin{center}
    \includegraphics[width=\linewidth]{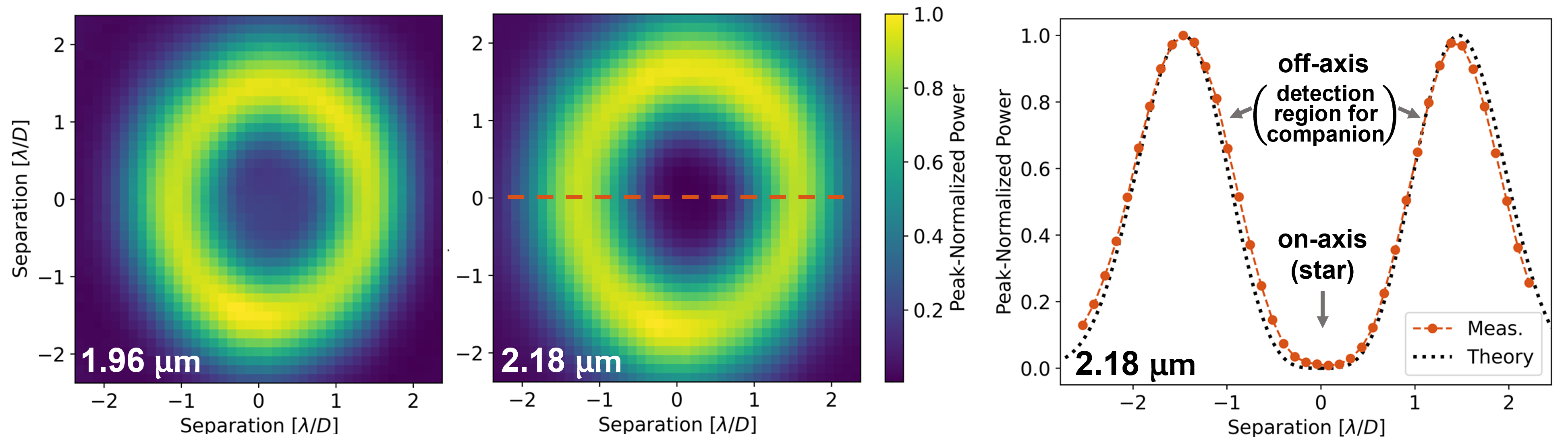}
    \end{center}
    \caption{
    Coupling maps of the KPIC VFN mode as measured at the telescope using an internal calibration source. The 2D maps show the power coupled into the fiber as a function of the image position measured using NIRSPEC at 1.96~$\mu$m (left) and 2.18~$\mu$m (middle). The orange dot-dashed curve in the right plot is a horizontal crosscut through the middle of the 2.18~$\mu$m coupling map, as indicated by the dashed orange line in the 2D map. The black dotted line in the right plot is the theoretical expected performance. The center region, with low power, is where the star would be while the sides, with higher coupled power, are where the companion would be. All plots and curves are normalized to the peak value in their corresponding data.
    \label{fig:DaytimeNIRSPEC}} 
\end{figure}

Once installed at the telescope, the KPIC VFN mode was tested using an internal light source to validate the system performance. An SMF was inserted at the focal plane input of the FIU, indicated by the converging point in the left panel of Fig.~\ref{fig:OpticalLayout}, to simulate a diffraction-limited focused beam received from the AO system. The FAM was then used to scan the PSF over the fiber to sample the coupling versus the image position, thereby generating a 2D map of the power coupled into the fiber from each point in the VFN field-of-view (FOV). Using NIRSPEC as the detector, this scan provided a spectrally-resolved coupling map. Figure~\ref{fig:DaytimeNIRSPEC} shows the average coupled power in the shortest and middle K band echelle orders on NIRSPEC (centered at 1.96~$\mu$m and 2.18~$\mu$m). The center region of the 1.96~$\mu$m coupling map has more power because it is further from the design wavelength of the \ed{narrowband} vortex \ed{in use} and hence has higher leakage and a worse null. The bright region for both coupling maps is relatively constant regardless of azimuthal angle, and the \ed{orange dot-dashed} crosscut in the rightmost plot shows the symmetry across two sides of the map. The high symmetry is an indicator that the WFE in the system as corrected for this test was very low, as WFE leads to a shift in power around the bright fringe. 
There is a slight vertical elongation in the maps which is due to an asymmetric response in the FAM axes from the way this scan was performed and is not a real effect from VFN; the on-sky coupling region is symmetric but has a gentle 6-lobed modulation due to the Keck primary mirror's hexagonal shape\cite{Ruane2019SPIE}. \ed{The crosscut is in nice agreement with the black dotted line that is a two-sided, peak-normalized replica of the orange charge 2 curve from Fig.~\ref{fig:Concept}(c) showing the theoretically expected performance.} This validates the fact that VFN has a near-circularly-symmetric bright fringe capable of simultaneously searching a full annular region around a star.

\section{On-Sky Commissioning} \label{sec:data}
The VFN mode was installed on KPIC alongside the other Phase II upgrades in February 2022. On-sky commissioning began shortly thereafter, 
with observations performed on UT 14 April 2022, 14 November 2022, and 6 January 2023. A total of 5 stars were observed for this dataset, all of them known to not have companions within the VFN FOV and to have a small enough angular diameter such that they could be considered point sources ($<$0.03~$\lambda/D$). To test for known multiplicity, the stars were cross-referenced against the Ninth Catalog of Spectroscopic Binaries (SB9)\cite{Pourbaix2004_SB9} as well as the Washington Double Star Catalog (WDS)\cite{Mason2021_WDS}. None of the targets were present in the SB9 and only one, 79 Cyg, was present in the WDS though its closest companion entry is at $1.6^{\prime\prime}$ (${>}35~\lambda/D$) and hence far out of the VFN FOV. In addition, both HD~213179 and 79~Cyg are present in the catalog of calibrator stars for interferometers\cite{Swihart2017_InterferometryCatalog} such that they have been further pre-vetted for use in interferometric observations. The angular diameters were obtained from the Mid-infrared stellar Diameters and Fluxes (MDF) Catalog\cite{Cruzalebes2019_MDFC}. 
  

\subsection{Measurables} \label{sec:measurables}
The measurable performance metric for KPIC is the counts per spectral channel on NIRSPEC, from which the end-to-end (E2E) throughput of the instrument can be computed. This throughput, encompassing all losses in the optical path, is different than the coupling efficiency, $\eta$, defined in Eq.~\ref{eq:overlap}. Coupling efficiency is the power coupled into the fiber normalized by the power incident on the fiber and hence only considers losses due to modal mismatch between the SMF fundamental mode and the system PSF. In the laboratory, we have been able to measure the coupling efficiency directly by measuring the flux incident on the fiber tip with a power meter and comparing to the power at the immediate output of the fiber\cite{Echeverri2019_VFN,Echeverri2021_BroadbandVFN}. KPIC, however, does not currently have a way of sampling the power incident on the fiber or immediately after the fiber so the coupling efficiency cannot be directly determined; NIRSPEC flux and E2E throughput are the closest accessible measures. As shown in Sec.~\ref{sec:SNR}, the E2E throughput is sufficient for computing the expected SNR for a given observation.

The E2E throughput, $T$, is computed by comparing the counts on NIRSPEC to the expected flux for the given star. This is done with the throughput calculator in the KPIC Data Reduction Pipeline.\footnote{\href{https://github.com/kpicteam/kpic\_pipeline}{https://github.com/kpicteam/kpic\_pipeline}}
The star's effective temperature is first used to generate a blackbody curve which is scaled by the star's apparent $K$ band magnitude and the collecting area of the Keck primary mirror (76~m$^2$). The measured counts on NIRSPEC for a given frame are converted to flux assuming a detector gain of $g$=3.03~[$\mathrm{e^-/ADU}$]~\cite{Lopez2020_NIRSPECUpgradeSpecs} and accounting for the frame integration time. The ratio of the two fluxes is then computed at each spectral channel and the end result is a wavelength-dependent throughput measurement, $T(\lambda)$ which includes all losses in the optical path from the top of the atmosphere to the detector including atmospheric transmission, optical coatings, coupling efficiency, quantum efficiency, and more. 
The detected signal, in counts [ADU], for an object can then be computed as 
\begin{equation} \label{eq:signal}
    F(\lambda) = \frac{T(\lambda)\Phi(\lambda)_{object}A\tau}{g}, 
\end{equation}
where $\Phi$ is the photometric flux for the object, $A$ is the collecting area, $\tau$ is the integration time, and $g$ is the detector gain. In this formulation, $T$ includes the quantum efficiency of the detector and so has units of [$\mathrm{e^-/ph}$]. $\Phi$ is the photon flux in the desired spectral bandwidth, $\Delta\lambda$, of the spectral channel such that it has units of [$\mathrm{ph/s/m^2}$]. We note that in this paper, we define a spectral channel as a single column of pixels perpendicular to the axis of dispersion. 
The re-imaged fiber results in a Gaussian-like PSF in this direction such that $\sim$3 pixels in the column are combined into a channel. The line-spread function of the NIRSPEC slit is oversampled to $\sim$3 pixels as well such that 3 spectral channels are combined to form a single spectral resolution element. Given the $R{\sim}$35,000 resolution of KPIC in K band, and the 3-pixel coverage, a spectral channel, as defined in this paper, thus subtends approximately $\Delta\lambda{=}2.1{\times}10^{-5}$~$\mu$m in the spectrum.

Due to the dependence of the coupling efficiency on angular separation, as shown in Fig.~\ref{fig:DaytimeNIRSPEC}, the throughput is also separation dependent such that it would more properly be expressed as $T(\lambda,\alpha)$, where $\alpha$ is the on-sky angular separation. However, for simplicity, we will refer to the $\alpha{=}0$ throughput as ``on-axis", and other throughput measurements as ``off-axis" with a specification for the separation. For SNR calculations, it is necessary to know $T$ for both the planet and the star. The star throughput in the VFN mode is measured by targeting a known-single star and observing it centered on the fiber through the vortex to provide the on-axis value. The planet throughput is measured by offsetting the star over the fiber so that it acts as an off-axis point source in the same way that the planet would at the offset separation.

\subsection{Observations}
For VFN commissioning, several throughput measurements were made on 5 stars over 3 separate nights. The target stars, their key properties, the nights on which they were observed, and the observing conditions are listed in Table~\ref{tab:OnAxData}. 

\begin{table}
    \def\arraystretch{1.5}
    \caption{Observations for VFN Commissioning}
    \label{tab:OnAxData}
    \small
    \begin{center} 
        \begin{tabular}{ccccccccc}
        \hline\hline
        \multirow{2}{*}{\makecell{Date \\ (UT)}} & \multirow{2}{*}{Star} & \multirow{2}{*}{\makecell{Seeing \\ (DIMM)}} & \multirow{2}{*}{\makecell{Mag. \\ (K band)}} & \multirow{2}{*}{\makecell{Spectral \\ Type}} & \multirow{2}{*}{\makecell{Diam. \\ (mas)}} & \multirow{2}{*}{\makecell{Elevation \\ ($^\circ$)}} & \multicolumn{2}{c}{\makecell{On-Axis \\ Throughput (\%)}} \\
         & & & & & & & VFN & DS \\ \hline\hline
        14 Apr 2022 & HIP 62944 & 1" & 4.12 & K3III & 0.71 & 75 & 0.07 & 2.15\\ \hline
        \multirow{2}{*}{14 Nov 2022} & HD 213179 & 0.6" & 2.98 & K2II & 1.27 & 63 & 0.11 & 3.23\\
         & 79 Cyg & 0.6" & 5.66 & A0V & 0.24 & 71 & 0.06 & 1.72\\ \hline
        \multirow{2}{*}{6 Jan 2023} & HIP 14719 & N/A & 6.32 & A0V & 0.18 & 70 & 0.07 & 2.55 \\
         & HIP 18717 & N/A & 6.07 & A0V & 0.21 & 83 & 0.06 & 2.32 \\ \hline
        \end{tabular}
    \end{center}
    Seeing from Maunakea Weather Center seeing monitors. Conditions were unavailable for 6 Jan 2023. 
    \\ 
    Magnitude, spectral type, and diameter from the MDF Catalog~\cite{Cruzalebes2019_MDFC}.\\
    Elevation is the average elevation of the target during the observations.\\
    On-axis throughput is the average in echelle order 4 with the star aligned to the center of the SMF.\\ 
\end{table}

\if false
    \begin{table}
        \def\arraystretch{1.5}
        \caption{On-axis Data Collected During Commissioning in VFN and DS Mode}
        \label{tab:OnAxData}
        \begin{center}
        \begin{tabular}{ccccccc}
        \hline\hline
        \begin{tabular}[c]{@{}c@{}}Date\\ (UT)\end{tabular} & Star & \begin{tabular}[c]{@{}c@{}}Seeing\\ (DIMM)\end{tabular} & \begin{tabular}[c]{@{}c@{}}Magnitude\\ (K band)\end{tabular} & \begin{tabular}[c]{@{}c@{}}Diameter\\ (mas)\end{tabular} & \begin{tabular}[c]{@{}c@{}}Elevation\\ ($^\circ$)\end{tabular} & \begin{tabular}[c]{@{}c@{}}Star Throughput \\ (\%)\end{tabular} \\ \hline\hline
        14 Apr 2022 & HIP 62944 & 1" & 4.12 & 0.71 & 75 & 0.07 \\ \hline
        \multirow{2}{*}{14 Nov 2022} & HD 213179 & 0.6" & 2.98 & 1.27 & 63 & 0.11 \\
         & 79 Cyg & 0.6" & 5.66 & 0.24 & 71 & 0.06 \\ \hline
        \multirow{2}{*}{6 Jan 2023} & HIP 14719 & N/A & 6.32 & 0.18 & 70 & 0.07 \\
         & HIP 18717 & N/A & 6.07 & 0.21 & 83 & 0.06 \\ \hline
        \end{tabular}
        \end{center}
        Seeing conditions from Maunakea Weather Center seeing monitors. Conditions for 6 Jan 2023 were unavailable
        .\\ 
        Magnitude and diameter from the MDF Catalog~\cite{Cruzalebes2019_MDFC}.\\
        Elevation is the average elevation of the target during the observations.\\
        Reported throughput is the average in echelle order 4 of NIRSPEC.\\ 
    \end{table}
\fi

\begin{figure}
    \begin{center}
    \includegraphics[width=\linewidth]{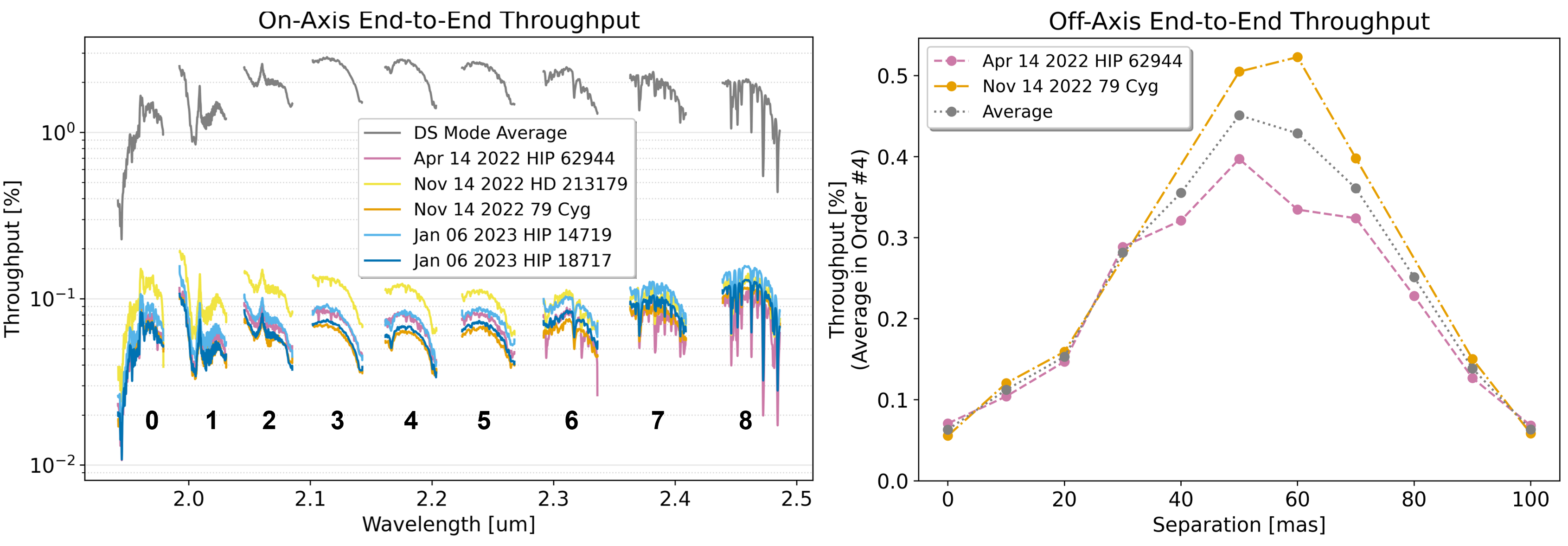}
    \end{center}
    \caption{
    Left: End-to-end on-axis throughput. The grey line shows the average DS mode throughput for all of the nights. The other lines are the VFN mode throughput measurements, where each line is the median of multiple samples on the given target. The spectra have been downsampled to R=3,500 to smooth out spectral lines for a simpler view. The VFN mode measurements represent the throughput for the host star during a VFN observation. Right: Average end-to-end throughput in echelle order 4 (2.16-2.20~$\mu$m) for points at varying separations. This represents the companion throughput given the off-axis separation from the host star. The orange dot-dashed line is for the data on 79 Cyg, the pink dashed line is for the HIP 62944 data, and the gray dotted line is the average of the two. Note that the y-axis is re-scaled compared to the on-axis plot.
    \label{fig:Throughputs}} 
\end{figure}

The on-axis throughput was measured with multiple samples per star to account for variability in seeing and turbulence within individual NIRSPEC frames. The left panel of Fig.~\ref{fig:Throughputs} shows the on-axis throughput in the DS mode (no vortex; grey) and VFN mode (with vortex; all other colors). The DS mode line is the average of the DS mode throughput measurements for all the targets. The VFN mode lines are the median of the various samples for each target. Note that the spectra are downsampled from the true spectral resolution of 35,000 to 3,500 in this figure for visualization purposes. 
The $\sim20$~nm gaps along the wavelength axis are due to the format of the two-dimensional spectra generated by the cross-disperser on the NIRSPEC detector. The resulting windows of signal are the echelle orders and are numbered in KPIC from zero to eight from left to right (short to long wavelengths). Note that KPIC orders 0 to 8 correspond to NIRSPEC orders 39 to 31, respectively, following the conventional NIRSPEC numbering; we renumber them in KPIC for simplicity. 
The average on-axis throughput in echelle order 4, the middle order on NIRSPEC, is included as the rightmost columns in Table~\ref{tab:OnAxData} for both modes on each star. The VFN mode on-axis measurements represent the throughput for the host star in a VFN observation since the star is aligned to the center of the fiber. 

The off-axis throughput was measured by scanning the stellar PSF along a radial line away from the center of the fiber. This samples various separations so that a line profile of throughput versus separation can be generated. The throughput for a companion at a given separation can then be determined from these curves. Line scans were only performed on two of the targets due to time constraints. The right panel of Fig.~\ref{fig:Throughputs} shows the average throughput in echelle order 4 for these two line scans. The orange dot-dashed curve from the November data on 79 Cyg is a single radial scan while the pink dashed curve from April on HIP 62944 is the average of two scans on opposite sides of the star. The gray dotted curve is the average of the 79 Cyg and and HIP 62944 data to provide a representative off-axis throughput for the KPIC VFN mode.

\section{Analysis} \label{sec:analysis}
Figure~\ref{fig:Throughputs} shows that the VFN on-axis throughput achieved on-sky is consistent across multiple targets and over nine months. All the VFN measurements are over an order of magnitude lower than the DS mode (grey) throughput, showing that the VFN mode is successfully rejecting the on-axis starlight. 
The rightmost columns in Table~\ref{tab:OnAxData} show that the VFN mode consistently reduces the on-axis throughput in echelle order 4 by a factor of ${\sim}3{\times}10^{-2}$ relative to the DS mode, with the more recent January data showing the best relative reduction.  The HD 213179 measurement seems to be a slight outlier with the highest VFN on-axis star throughput in Fig.~\ref{fig:Throughputs}, but it is still within a factor of 2 from the lowest throughput measurement, which was achieved on 79 Cyg that same night. The other four on-axis measurements are all within a factor of 1.3 from each other with the average for each star in echelle order 4 varying between 0.056\% and 0.073\%. Improvements in starlight rejection may be achieved through the work described in Section~\ref{sec:NextSteps} but we see that the system can already be reliably calibrated and controlled to the same level over months and the performance is stable and predictable. 

\ed{
As described in the beginning of Sec.~\ref{sec:measurables}, KPIC does not have a way to measure the coupling efficiency, $\eta$, defined in Sec.~\ref{sec:VFN}. However, we can approximate the VFN coupling using the throughput ratio to the DS mode and making assumptions about the DS performance based on knowledge from KPIC Phase I. The on-axis ratio is
\begin{displaymath}
    \frac{T_{VFN}}{T_{DS}}\approx\frac{\eta_{VFN}}{\eta_{DS,ap}\mathrm{SR}},
\end{displaymath}
where $\eta_{VFN}$ is the VFN coupling efficiency, $\eta_{DS,ap}$ is the coupling efficiency term for the DS mode set by the aperture shape, and $\mathrm{SR}$ is the Strehl ratio in DS mode which has a direct impact on coupling efficiency~\cite{Jovanovic2017_SMFOnSky}. Other throughput terms (transmission of optics, quantum efficiency, etc.) cancel out since they are equal for both modes. For the Keck aperture, $\eta_{DS,ap}$ is at most 67\%, assuming no static WFE in the telescope or instrument, and $\mathrm{SR}{\sim}55\%$ as set by the SHWFS for stars with the brightness in our dataset~\cite{Wizinowich2015_K2AOPerformance}. Note that by the Mar\'echal approximation~\cite{Ross2009_SR2RMSWFE}, 55\% corresponds to ${\sim}250$~nm RMS of wavefront error. Taking the HIP 18717 data, which shows the best VFN starlight rejection, we get $T_{VFN}=0.06\%$ and $T_{DS}=2.32\%$, which provides $\eta_{VFN}{\approx}9.5\times10^{-3}$. In 2019, we predicted that the KPIC VFN mode would achieve $6\times10^{-3}$ by using KPIC Phase I wavefront measurements from the PyWFS~\cite{Echeverri2019b_VFN} which, as described in Sec.~\ref{sec:ImproveStar}, has smaller residuals than the SHWFS. In those predictions, the residual WFE was the key limiting term for the charge 2 vortex.
}

\ed{
This estimate of the on-axis VFN coupling efficiency shows that the on-sky null roughly matches expectation though there is some discrepancy. The estimate makes several assumptions that need to be further evaluated before a precise on-axis coupling efficiency can be reported. For example, $\eta_{DS,ap}$ and $\mathrm{SR}$ may be lower than assumed, the power spectrum of the WF residuals may be different with the SHWFS and hence might include more low-order WFE that VFN is sensitive to, and there may be additional unaccounted-for losses such as a smaller aperture size in the VFN mode, among other inaccuracies in the estimate. We will rely on throughput for the remainder of the paper since it is measurable and well-determined in KPIC as described earlier.
}

The average off-axis throughput in echelle order 4 (gray curve in right plot of Fig.~\ref{fig:Throughputs}) shows a peak of 0.45\% at 50~mas, with the true maximum likely occurring somewhere between the 50 and 60~mas samples given the shape of the curve. From Fig.~\ref{fig:Concept}(c), the peak is expected to occur at around 1.4$\lambda/D$ which at 2.2~$\mu$m on the Keck Telescope would be ${\sim}60$~mas. Thus, the location of the maximum is in close agreement with the predicted performance, especially for the 79 Cyg line scan. The line scan data provides the throughput for an off-axis point source meaning that KPIC VFN would obtain 0.45\% of the total light from a companion at 50~mas from its host star. 

We can contextualize the VFN companion throughput by comparing to the KPIC DS mode. In the DS mode, the analogous throughput is provided by the on-axis value since the SMF is aligned to the companion. The rightmost column in Table~\ref{tab:OnAxData} shows that on the two stars for which we took VFN line scans, the DS mode achieved an average on-axis throughput of 2.15\% and 1.72\%. The off-axis peak in the VFN mode was 0.40\% and 0.52\%, respectively, as shown in the right plot of Fig.~\ref{fig:Throughputs}. Thus, the end-to-end throughput for the companion in the VFN mode is 20 to 30\% of that in the DS mode. This is a key result to highlight. The relative VFN throughput is higher than intuited from the ${\sim}$10\% peak in Fig.~\ref{fig:Concept}(c) because of the distinction made between coupling efficiency and throughput at the beginning of Sec.~\ref{sec:measurables}. Coupling efficiency, and hence Fig.~\ref{fig:Concept}, consider the coupled power relative to the total power incident on the fiber whereas here we are considering the relative throughput for VFN versus DS mode. The DS mode does not couple all the light incident on the SMF either; the maximum DS coupling on the Keck aperture is 67\%\cite{Ruane2019SPIE}. Thus, in the absence of WFE, VFN achieves 16.5\% of the DS mode coupling. Additionally, the DS mode is more sensitive to WFE than the VFN mode. In the VFN mode, low-order aberrations simply shift power around the bright fringe but average over time to no net loss since the throughput can be instantaneously increased or decreased for a given off-axis position. In the DS mode, all aberrations strictly decrease the on-axis (companion) throughput. We refer readers to the lower-left panel of Fig.~7a in Echeverri et al.~\cite{Echeverri2019b_VFN} which shows that in the presence of realistic on-sky WFE, the VFN coupling only drops from 11\% to 8\% and remains relatively constant at that value as the WFE varies. 

We note that the majority of the losses in the instrument, in both VFN and DS mode, are due to the number of optics in the optical path, not the use of single mode fibers. Delorme et al.~\cite{Delorme2021_KPIC} demonstrate this with a detailed accounting of losses in the KPIC Phase I system. 

\ed{We can further contextualize the VFN performance more generally by comparing to an unresolved (ie. seeing-limited) observation without a fiber. In this unresolved case, there is no starlight suppression so both the companion and star have the same instrument throughput. In the VFN case, the companion throughput is approximately 6.9 times higher than the star throughput, given the system performance presented above. This increases the effective contrast relative to the star by the same amount, allowing the VFN mode to target fainter companions. The next section further explores the VFN performance and predicted SNR limits.}

\if False
    NOTE: Removing the DS mode leaves out several elements that need to be considered:
    \begin{itemize}
        \item Loses rationale for how to select the best performance to use as "current VFN performance" in SNR section. In first paper version, used the one that provides the best VFN/DS null. This allowed us to ensure that the null isn't deep because of poor coupling/throughput that day in general (imagine strong cloud coverage - VFN null throughput would seem very good/deep but the DS mode would show that it's actually because throughput that day was poor anyway). Without this normalization, we need to find new rationale for how to select values for use in SNR section.
        \item Loses good VFN mode planet throughput. When we were dividing by the DS mode throughput, it was clear that the Nov 14 79Cyg line scan had very high throughput relative to the DS mode. This ultimately led to an averaged E2E planet throughput of .75\% when rescaled. Now our averaged planet throughput will likely be closer to 0.45\% which is a big drop. This drop won't be compensated-for by the drop in null. Need to think about implications of this.
        \item \textbf{Combination of the last two points:} Since we don't have a night with good null and line scan on the same star, we'll need to combine data from separate nights. The variability in conditions between nights will no longer be normalized out by the DS mode so we need to think about the validity of how we combine the data. - could maybe use averages?
        \item Loses ability to re-scale to more typical nights. As such, the VFN planet throughput will be lower than typical since the VFN nights were generally rougher conditions. 
        \item Removing the DS-mode measurement removes the direct comparison of VFN-mode planet throughput to DS mode planet throughput (since we lose right panel of Fig. 4 on original version). As such, we can't readily make the comment/argument about how VFN mode actually has a non-intuitively high throughput due to DS mode losses. - Could maybe bring that comment back in by saying "VFN mode achieved approx 0.45\% throughput on companion. For reference, DS mode achieves a peak of 3\% on typical nights s.t. VFN is achieving approx 15\% of DS mode... this is higher than the 10\% because of DS-mode losses, etc.". However, this 15\% is much lower than the 25\% we have when we consider VFN/DS ratio directly  from the same night... Could maybe bring that ratio in for this part only. Or, rather than using 3\% as the typical value, could say, ``on the two nights where we did line-scans, the peak DS mode throughput was XX\% such that the VFN mode was achieving YY\% of what the DS mode was hitting. This is unintuitively high since..."    
    \end{itemize}
\fi

\subsection{SNR Formulas} \label{sec:SNR}
The performance of an observation can be predicted by using the end-to-end throughput values to determine the expected signal on the detector and from there, computing the SNR for the companion. The SNR per spectral channel, including typical noise sources, is 
\begin{displaymath}
    \mathrm{SNR}=\frac{S_p}{\sqrt{S_s + S_p + S_b + \mathrm{RN}}},
\end{displaymath}
where $S_p$ and $S_s$ are the planet and star signal per Eq.~\ref{eq:signal} but without the gain term to maintain units of [$\mathrm{e^-/s}$], $S_b$ is the thermal background signal, and RN is the read noise \ed{squared} per frame on the detector. As mentioned in Sec.~\ref{sec:obsStrat}, the background signal in KPIC data is primarily dominated by instrumental background inside NIRSPEC and can be expressed as $S_b=B_cn_{pix}\tau$, where $n_{pix}{=}3$ is the number of pixels combined for a spectral channel and $B_c$ is the number of background electrons received per second. In Phase II of KPIC, $B_c$ is wavelength dependent but is generally $\sim$1~[$\mathrm{e^-/pix/s}$]. The read noise accumulates on a per-frame basis and can be expressed as $\mathrm{RN}=R_c^2n_{pix}n_{frame}$ where $R_c$ was measured at $\sim$10~[$\mathrm{e^-/pix/frame}$] by Lopez et al.~\cite{Lopez2020_NIRSPECUpgradeSpecs}  

Given the linear range of 25,000~ADU per pixel for NIRSPEC~\cite{Lopez2020_NIRSPECUpgradeSpecs}, the residual VFN starlight throughput, and the electron rate of the background, the exposure time of individual frames can be set to make read noise irrelevant for KPIC VFN observations. For faint targets, long exposures can be acquired such that the background signal will outpace the read noise and the observation will be primarily background-limited. For bright targets, short exposures will be needed to avoid saturating on the starlight signal but the star will thus outpace the read noise such that the observation becomes stellar-photon-noise-limited. 

The read noise will therefore be omitted from further SNR calculations in this work without loss of generality given that the frame exposure time for an observation is set to ensure that other terms dominate. The SNR equation now simplifies to
\begin{equation} \label{eq:SNR_full}
    \mathrm{SNR}=\frac{T_p\epsilon\Phi_s A\sqrt{\tau}}{\sqrt{\Phi_s A (T_s+\epsilon T_p) + B_c n_{pix}}},
\end{equation}
where the full expressions for the stellar, planet, and background signals have been filled in. $T_s$ and $T_p$ are the star (on-axis) and planet (off-axis) throughput respectively, $\Phi_s$ is the flux of the star, and $\epsilon$ is the photometric flux ratio between the planet and the star, $\Phi_p/\Phi_s$. Solving for $\tau$ in Eq.~\ref{eq:SNR_full}, the total integration time required to achieve a goal SNR per spectral channel can then be expressed as
\begin{equation} \label{eq:itimeForSNR_full}
    \tau=\left( \frac{\mathrm{SNR}}{T_p\epsilon\Phi_sA} \right)^2 \left( \Phi_sA (T_s+\epsilon T_p) + B_cn_{pix} \right).
\end{equation}

\ed{
For bright stars, a further simplification can be made by omitting the background noise and assuming observations are stellar-photon-noise-limited. The background noise will remain one tenth of the stellar photon noise, and hence regarded as negligible, for
\begin{displaymath}
    \frac{S_s}{S_b} = 10 = \frac{T_s\Phi_sA}{B_cn_{pix}}.
\end{displaymath}
This leads to a stellar K band magnitude of 6 below which the contribution of the background photon noise to the SNR can be ignored in favor of the dominating stellar photon noise. At a magnitude of 8, the background noise becomes equivalent to the stellar noise and beyond that, it starts to dominate. 
Equation~\ref{eq:itimeForSNR_full} thus simplifies to
\begin{equation} \label{eq:itimeForSNR_SPNL}
    \tau=\frac{T_s}{T_p^2}\frac{\mathrm{SNR}^2}{\epsilon^2\Phi_sA} ,
\end{equation}
in the stellar-photon-noise-limited regime.
}

\subsection{Predicted Performance and Detection Limits} \label{sec:detectLimits}
Using Eq.~\ref{eq:itimeForSNR_full} and the throughput values presented above, we can predict the KPIC VFN capabilities from the on-sky commissioning performance. Given that the typical VFN on-axis throughput in echelle order 4 is between 0.056\% and 0.073\%, a value of $T_s{=}$0.065\% will be assumed for this section. The average line scan performance (gray dotted curve in the right plot of Fig.~\ref{fig:Throughputs}) will be used for $T_p(\alpha)$ with a peak value of 0.45\% at $\alpha{=}$50~mas and interpolating on the curve for other separations as needed. To define a goal SNR value, we consider the abundance of science data acquired with KPIC throughout Phase I and Phase II, and find that the instrument can detect a companion at a cross-correlation function (CCF) SNR of 3-5 when the SNR per spectral channel is around 1. This boost between CCF SNR and SNR per channel is primarily guided by the number of lines in the spectra\cite{Snellen2015_HDC,Wang2017_HDC} and is currently limited by the ability to account for systematics such as spectral fringing. Thus, for this work, we will only consider a companion ``detected" when a goal SNR per spectral channel of 3 has been met to ensure an unambiguous detection. 

\begin{figure}
    \begin{center}
    \includegraphics[width=0.6\linewidth]{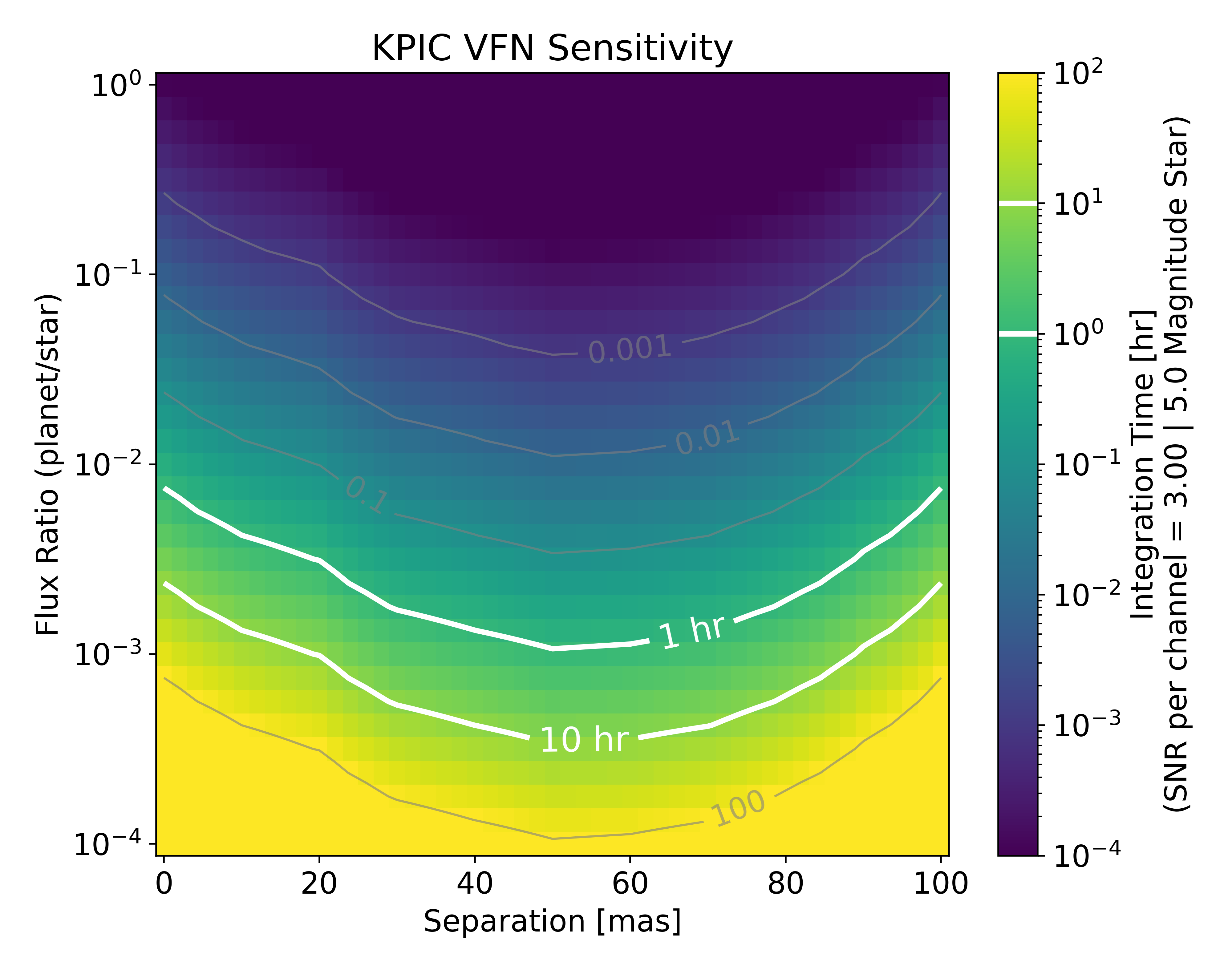}
    \end{center}
    \caption{
    Integration time required to reach an SNR per spectral channel of 3.0 on a companion at a given flux ratio and separation around a star with a K band magnitude of 5. This assumes the current KPIC VFN charge 2 performance demonstrated on-sky in this paper. The white lines show the 1 hour and 10 hour contours while light grey lines show other powers of 10. This includes planet and star photon noise along with background noise, but neglects read noise which is negligible. For a 5\textsuperscript{th} magnitude star as shown here, stellar photon noise dominates.
    \label{fig:SNRPlot}} 
\end{figure}

\begin{table}
    \caption{Values for SNR calculation in Fig.~\ref{fig:SNRPlot}}
    \label{tab:SNRValues}
    \begin{center}
    \begin{tabular}{l|l}
        \hline\hline
        \multicolumn{2}{c}{Observing Parameters}                            \\ \hline\hline
        Star Throughput ($T_s$)              & 0.065\%                      \\
        Peak Companion Throughput ($T_p$)    & 0.45\%                       \\
        Star K Band Magnitude                & 5.0                          \\
        Central Wavelength ($\lambda$)       & 2.18~$\mu$m                  \\
        Channel Width ($\Delta\lambda$)      & $2.1\times10^{-5}$~$\mu$m    \\
        Photon Rate \ed{for a 5\textsuperscript{th} mag. Star}  ($\Phi_s$)          
                                             & 1,000~ph/s/m$^2$             \\ 
        Goal SNR per Channel                             & 3.0                          \\ \hline\hline
        \multicolumn{2}{c}{Instrument Constants}                            \\ \hline\hline
        Collecting Area ($A$)                & 76~m$^2$                     \\
        Background Electron Rate ($B_c$)     & 1~e$^-$/pix/s                \\
        Pixels Per Channel ($n_{pix}$)       & 3                            \\ \hline\hline
    \end{tabular}
    \end{center}
\end{table}

With these values, Fig.~\ref{fig:SNRPlot} shows the integration time required to reach an SNR per channel of 3 on a companion at a varying flux ratio ($\epsilon$) and separation ($\alpha$) from a host star with a K band magnitude of 5. The white contour lines show the 1 and 10 hour detection limits. Table~\ref{tab:SNRValues} summarizes the values used for these calculations. The results show that, given the current instrument performance, the KPIC VFN mode should be able to detect companions within 1 hour down to a flux ratio of $10^{-3}$ around stars brighter than fifth magnitude. 

On fainter stars, KPIC VFN will start to become background-noise limited due to the NIRSPEC instrument background (see Sec.~\ref{sec:obsStrat}) but can still detect companions several hundred times fainter than their host star. For example, the current performance enables the instrument to detect a system like GQ~Lup~AB\cite{Neuhauser2005_GQLup} in under 1 hour if the brown dwarf companion ($K{=}13$) was located at 50~mas from its host star ($K{=}7.1$). 

Note that this result uses the VFN performance in echelle order 4 ($\sim$2.15-2.20~$\mu$m), which is at the center of the K band window covered by KPIC. However, the CO bandhead, which contains the bulk of the spectral lines used for detection on KPIC, starts at $\sim$2.29~$\mu$m and extends longwards such that it subtends echelle orders 6 to 8. Thus, these are the key orders in which higher SNR is most valuable, even if they are not where the current monochromatic charge 2 vortex mask used by the KPIC VFN mode achieves the deepest starlight rejection. The left plot in Fig.~\ref{fig:Throughputs} shows that the starlight rejection in echelle order 8 is approximately two times worse than order 4. This leads to a correction factor of $\sim$2 in the required integration time if only order 8 were considered. Realistically, a detection combines multiple orders such that the integration time will likely be somewhere between the order 4 and order 8 values.

A key point to make is that Fig.~\ref{fig:SNRPlot} is valid for a companion at a given separation regardless of the on-sky position angle since the VFN coupling is axially symmetric. This makes VFN a prime technique for new detection campaigns. Other techniques may achieve a higher SNR in less time for a known companion when they can be aligned directly to maximize throughput on the target. However, those techniques cannot simultaneously cover the full region that VFN covers, so the integration time required to search the same area as VFN with equivalent SNR is significantly higher. 


\section{Next Steps} \label{sec:NextSteps}
The current KPIC VFN performance is promising and consistent over months, and demonstrates that the instrument mode is ready for on-sky operation in surveys searching for new companions. There is, nevertheless, still room for further improvement through on-going work on both the hardware and software. These efforts primarily target either improving the starlight rejection (smaller $T_s$) or increasing the planet throughput (larger $T_p$). 

\subsection{Improving Starlight Rejection} \label{sec:ImproveStar}

With a charge 2 vortex mask, like the one currently installed in the instrument, the largest source of coupled starlight is due to wavefront aberrations. This is because charge 2 is significantly less sensitive to tip-tilt jitter, differential atmospheric refraction (DAR), and the finite angular size of the star~\cite{Ruane2019SPIE}. The current tip-tilt residuals of $\sim$6-7~mas~RMS put jitter at the next largest source of starlight but likely an order of magnitude or more below the starlight from WFE, making improved wavefront control the best way to reduce the coupled starlight.  

One route for accomplishing this is to switch from using the SHWFS to the PyWFS. 
The latter has an improved sensitivity to low order wavefront aberrations, which are what VFN is most sensitive to given its small working angle. The PyWFS has demonstrated an improvement, over the SHWFS, of a factor of 2 in raw contrast at small separations (${\sim}2\lambda/D$) with the NIRC2 charge 2 vortex coronagraph~\cite{Bond2020_PyWFS}, which would correspond to a reduction of $\sqrt{2}$ in WFE. Furthermore, predictive control is being implemented with the PyWFS and preliminary results show a reduction by a factor of $\sim$1.2 in the median RMS wavefront residuals on-sky compared to the default integrator control~\cite{VanKooten2021_PyWFSPredictive}. As shown in Fig.~\ref{fig:OpticalLayout}, the PyWFS is already installed in the system and, in fact, it was the primary wavefront sensor for KPIC during Phase I. It is still being recommissioned for Phase II operations but requires only software modifications to support the new elements of the system. Once available for KPIC observations, it should provide a reduction by a factor of $>$1.5 in RMS WFE which in turn would yield a reduction of 2.25 in coupled starlight since the VFN starlight rejection is quadratically dependent on WFE~\cite{Ruane2019SPIE}. \ed{For targets with bright host stars where the observation is currently stellar photon-noise limited, this} leads to a \ed{corresponding factor of 2.25} reduction in integration time compared to the performance from Section~\ref{sec:analysis}. 
An additional reduction in WFE can also be expected from the increased temporal bandwidth provided by a new real time computer being installed in the summer of 2023 to upgrade the facility AO system. The improvement from this upgrade is harder to predict. 


\subsection{Improving Planet Throughput} \label{sec:ImprovePlanet}
An alternative to reducing the coupled starlight is to improve the planet throughput. As shown by Eq.~\ref{eq:itimeForSNR_full} \ed{and~\ref{eq:itimeForSNR_SPNL}}, the integration time is inversely proportional to the square of the planet throughput such that this is a more efficient and direct route for improvement. The current planet throughput is already close to the best that can be achieved with a charge 2 mask but, as explained in Section~\ref{sec:VFN}, a charge 1 mask has close to double the throughput. Charge 1 also has the added benefit of peaking in throughput at $0.9$ rather than $1.4\lambda/D$, allowing KPIC VFN to target companions at even smaller separations. As such, a charge 1 vortex mask will be added to the KPIC pupil stage as part of a service mission scheduled for the Winter of 2023. This promises to increase the planet throughput to $T_p{=}0.8$\%, a factor of 1.8 higher than its current value, and therefore reduce the required integration time to achieve the same SNR by a factor of 3.25. It is important to note that the charge 1 mask is no more sensitive to WFE than the charge 2 mask since they both have a similar quadratic dependence on specific, complementary aberrations~\cite{Ruane2019SPIE}. However, charge 1 is more sensitive to tip-tilt jitter and DAR, which would then become the limiting terms in the achievable starlight rejection, setting a new $T_s{\sim}0.2$\% limit. 

To have the starlight contribution from jitter with a charge 1 mask be below the contribution from the current wavefront residuals, jitter would need to be reduced by about a factor of 3 to below 2~mas~RMS. Switching to the PyWFS should help with the jitter residuals since the pyramid is able to sense and correct tip and tilt better than the SHWFS. Additionally, the tracking system described in Section~\ref{sec:tracking} is currently being overhauled to reduce software overheads and operate faster, thereby allowing it to target higher frequencies, including several resonances that add a significant amount of power to the jitter residuals. Recent off-sky testing has shown an increase in control frequency by more than a factor of two and the control algorithm is now being adjusted from a simple integrator to specifically notch the known resonances. Between these two modifications, we expect to reduce the tip-tilt jitter so that starlight leakage from it is below that from the current wavefront residuals. 

In addition to the jitter, the DAR will be a limiting factor with the charge 1 mask. 
An atmospheric dispersion compensator (ADC) has been designed specifically to meet the VFN requirements following the same methodology of Wang et al.~\cite{Wang2020_ADC} and will be installed in the system alongside the charge 1 vortex mask. This will correct the DAR to below 1~mas at the edges of the band so that the starlight leakage from DAR will be two orders of magnitude or more below that from residual on-sky WFE. 


\section{Conclusion}
This paper presented the first on-sky demonstration of VFN, a new interferometric nulling technique aimed at both detecting and characterizing companions at or within one diffraction beam width from their host star. This first VFN prototype worked as expected and was shown to enable the detection of companions $10^{3}$ times fainter than the central host in 1 hour at a separation of 50~mas. The new mode, available as part of the KPIC instrument, will be used to search for and characterize companions indirectly detected by Gaia and RV surveys. We have clearly identified areas for improvements such as better wavefront and jitter control for enhanced starlight suppression. We will also be replacing the existing charge 2 vortex mask with an optimized charge 1 vortex, close to doubling the effective off-axis throughput, and thereby reducing the required integration time by a factor of 3.25. With its simple design and implementation in fiber-fed direct imaging instruments, the demonstration of VFN on sky with KPIC is a key milestone towards future instruments such as Keck-HISPEC and TMT-MODHIS ~\cite{Mawet2022_HISPEC}, both of which have baselined VFN as a core mode. Fiber nulling is currently being considered for a possible implementation on the Habitable Worlds Observatory mission, enabling the detection and characterization of exoplanets at inner working angles substantially smaller than standard coronagraphs typically allow.

\subsection*{Disclosures}
The authors have no relevant financial interests and no potential conflicts of interest to disclose in this work. 

\subsection* {Acknowledgments}
D. Echeverri is supported by a NASA Future Investigators in NASA Earth and Space Science and Technology (FINESST) fellowship under award \#80NSSC19K1423. D. Echeverri also acknowledges support from the Keck Visiting Scholars Program (KVSP) to install the Phase II upgrades required for KPIC VFN. 

Funding for KPIC has been provided by the California Institute of Technology, the Jet Propulsion Laboratory, the Heising-Simons Foundation (grants \#2015-129, \#2017-318 and \#2019-1312), the Simons Foundation (through the Caltech Center for Comparative Planetary Evolution), and the NSF under grant AST-1611623.

\vspace{2mm}
The authors wish to recognize and acknowledge the very significant cultural role and reverence that the summit of Maunakea has always had within the indigenous Hawaiian community. We are most fortunate to have the opportunity to conduct observations from this mountain.

\subsection* {Code, Data, and Materials Availability} 
The code for reducing the spectra and calculating E2E throughput is freely available on GitHub in the \textit{kpic\_pipeline} repository at: 
\href{https://github.com/kpicteam/kpic\_pipeline}{https://github.com/kpicteam/kpic\_pipeline}. This work used the \textit{4514a56} commit from 5 Sep 2022. All KPIC data will be available via the Keck Observatory Archive (KOA) at \href{https://www2.keck.hawaii.edu/koa/public/koa.php}{https://www2.keck.hawaii.edu/koa/public/koa.php}.

\subsection* {Biography}
\textbf{Daniel Echeverri} is a graduate student at Caltech working with Professor Dimitri Mawet in the Exoplanet Technologies Lab. His research focuses on high contrast imaging of exoplanets and is primarily centered on the development of the new Vortex Fiber Nulling technique. He is also involved in the development and deployment of the Keck Planet Imager and Characterizer (KPIC) instrument at the Keck II Telescope.


\bibliography{Library}   
\bibliographystyle{spiejour}   

\listoffigures
\listoftables

\end{spacing}
\end{document}